\documentclass[twocolumn,preprintnumbers,aps,prb,amsmath,amssymb]{revtex4}
\usepackage{graphicx}
\usepackage{dcolumn}
\usepackage{bm}

\begin{document}

\newcommand{\vo}{VO$_x$}
\title{Growth and properties of strained VO$_{x}$ thin films with controlled stoichiometry}

\author{A. D. Rata$^{1}$, A. R. Chezan$^{2}$, and T. Hibma$^{1}$}
\address{$^{1}$Chemical Physics, Materials Science Centre, Rijksuniversiteit Groningen,
\\Nijenborgh 4, Groningen 9747 AG, The Netherlands}
\address{$^{2}$Nuclear Solid State Physics, Materials Science Centre, Rijksuniversiteit Groningen,
\\ Nijenborgh 4, Groningen 9747 AG, The Netherlands}
\author{M. W. Haverkort, and L. H. Tjeng}
\address{\textrm{II}. Physikalisches Insitut, Universit\"{a}t zu K\"{o}ln, Z\"{u}lpicher Str. 77, 50937
K\"{o}ln, Germany}
\author{ H. H. Hsieh, H.-J. Lin and C. T. Chen}
\address{Synchrotron Radiation Research Center, Hsinchu 30077, Taiwan}

\date{\today}

\begin{abstract}

We have succeeded in growing epitaxial films of rocksalt {\vo} on
MgO(001) substrates. The oxygen content as a function of oxygen
flux was determined using $^{18}$O$_{2}$-RBS and the vanadium
valence using XAS. The upper and lower stoichiometry limits found
are similar to the ones known for bulk material (0.8$<$x$<$1.3).
From the RHEED oscillation period a large number of vacancies for
both vanadium and oxygen were deduced, i.e. $\approx$$16\%$ for
stoichiometric VO. These numbers are, surprisingly, very similar
to those for bulk material and consequently quite
strain-insensitive. XAS measurements reveal that the vacancies
give rise to strong low symmetry ligand fields to be present. The
electrical conductivity of the films is much lower than the
conductivity of bulk samples which we attribute to a decrease in
the direct overlap between $t_{2g}$ orbitals in the coherently
strained layers. The temperature dependence of the conductivity is
consistent with a variable range hopping mechanism.
\end{abstract}
\maketitle


\section{Introduction}

One of the major challenges in transition-metal oxide thin film
research is to grow the material in single crystal form with
controlled oxygen stoichiometry. This is not a trivial task
especially when the transition metal ion has multiple valences and
the variation of the oxygen content leads to a rich and complex
phase diagram. Important examples for such materials include the
binary oxide systems like the vanadium and titanium oxides
\cite{goodenough63,bruckner63,tsuda,imada,rao98}. The most common
and convenient growth methods in thin film research like pulsed
laser deposition and sputtering techniques are not very suitable
here, because of the high substrate temperatures and consequently
also high oxygen pressures usually used, giving less opportunity
to fine tune the oxygen stoichiometry. These techniques are
obviously the method of choice for ternary and quarternary oxides,
in which one can vary the material properties by tuning one of the
constituent cation concentrations while keeping that of the oxygen
fixed. For the binary oxides, however, one has to resort to true
molecular beam epitaxy (MBE) techniques, in which one has to
carefully balance the oxygen flux with respect to the metal
evaporation rates in order to obtain good control and tuning of
the stoichiometry. The question here is whether one can obtain
thin films with physical properties that are systematic,
controllable and tunable.

Here we report on our research project to grow single crystal thin
films of vanadium monoxide (VO$_{x}$), as well as on the study of
the physical properties and the electronic structure of these
films. Bulk vanadium monoxide has many intriguing properties that
are closely related to the issue of stoichiometry. While at first
sight the crystal seems to have the very simple $fcc$ rocksalt
structure, it actually can have a very wide range of varying
oxygen concentrations: values of $x=0.8-1.3$ have been reported
for bulk VO$_x$. Even more remarkable is that there is a large
number of both cation and anion vacancies, even for $x=1$, in
which case the concentrations are as high as $15\%$ \cite{banus}.
Stoichiometric VO always remains disordered, whereas ordering of
the vacancies was only reported for $x$ between $1.2$ and $1.3$
\cite{morinaga}. A systematic investigation of the physical
properties of polycrystalline {\vo} has been carried out by Banus
et al. \cite{banus} in the early 70's. {\vo} shows an interesting
transition from a metallic to a semiconducting type of behavior at
$x$ close to $1$, and the magnetic susceptibility can be described
by a Curie-Weiss law, suggesting that the V $3d$ electrons
partially localize \cite{tsuda}. Several models have been proposed
to explain these properties, in which the role of electron
correlation effects, band formation and vacancies are discussed
\cite{goodenough,mott,cox,neckel}.

Only very little work on films of {\vo} has been done up to know.
Metal supported vanadium oxides were investigated by several
authors in connection with catalytic properties. Thin V oxide
layers deposited on Cu(${100}$)\cite{kishi1,kishi2}and
Ni($110$)\cite{kishi2a} were found to consist of a
V$_{2}$O$_{3}$-like oxide phase after room temperature oxidation.
VO$_{x}$-like islands growth was reported on
Rh($111$)\cite{hartmann}. Surnev {\it et al.} found a VO/VO$_{2}$
- like phase at low coverage on Pd(${111}$) using scanning
tunnelling microscopy (STM) \cite{surnev1,surnev2}. Thin films of
{\vo} have also been grown on TiO$_{2}$ substrates, because of the
enhanced catalytic activity shown by titania-supported vanadium
oxides \cite{henrich}. Sambi {\it et
al.}\cite{sambi1,sambi2,sambi3} studied locally ordered VO$_{x}$
films with a thickness up to $4$ ML grown on TiO$_{2}$(110) by
evaporating vanadium metal at room temperature and subsequent
annealing in vacuum for a short period of time, in order to
promote the uptake of oxygen from the substrate. From a structural
point of view, local order of their films is supported by X-ray
photoelectron diffraction (XPD).

In the examples mentioned above, the vanadium oxidation state was
established on the basis of the position and shape of the V
$2p_{3/2}$ photoemission line. However, the uncertainty of
stoichiometry of vanadia films remains, since the binding energies
(BE) of V $2p_{3/2}$ for different oxidation states are very
close, a comparison between data obtained in Refs.\cite{kishi1}
and \cite{sambi1} showing some discrepancies. Moreover, a
determination of the physical properties of {\vo} thin films is
missing in these studies. Preparation is often found to be
difficult since {\vo} easily oxidizes to V$_2$O$_3$. These
problems can be avoided by growing epitaxial films of {\vo} on
appropriate substrates with a careful optimization of the
evaporation conditions.

This paper describes how we have been able to successfully grow
coherent epitaxial single crystalline VO$_x$ thin films by
stabilizing them on MgO(100) substrates. The paper is organized as
follows. The details of the experimental conditions and equipment
used are given in the experimental section (section II). The
results are described in several subsections. We first present
(section IIIA) the growth process as monitored structurally
\textit{in-situ} by reflection high-energy electron diffraction
(RHEED) and low-energy electron diffraction (LEED) and
compositionally by x-ray photoelectron spectroscopy (XPS). We then
(section IIIB) determines the stoichiometry $x$ of the various
VO$_x$ films by using Rutherford backscattering spectrometry
(RBS). Further structural details of the epitaxy (section IIIC) is
revealed by \textit{ex-situ} x-ray diffraction (XRD). The
important issue of vacancy concentrations is addressed using a
combination of the RBS and RHEED intensity oscillations (section
IIID). Basic data about the electronic structure, such as the
valence and local crystal fields of the ions, (section IIIE) are
measured using soft-x-ray absorption spectroscopy (XAS). We end
the results section by presenting the temperature dependent
transport properties of the films (section IIIF). Finally we
discuss the results by comparing the properties of the VO$_x$ thin
films with those of the bulk material, and with the predictions
made by existing theoretical models .

\section{Experimental}

The experiments were performed in an ultra-high vacuum (UHV) MBE
system, with a base pressure below 1x10$^{-10}$ mbar, equipped
with RHEED, LEED and XPS facilities. RBS and XRD were used to
characterize the thin films grown in the MBE system {\it ex-situ}.
MgO(100) was chosen as the substrate because it has the same cubic
rocksalt structure as {\vo}, with a mismatch of about $3\%$. MgO
blocks were cleaved {\it ex-situ} and immediately introduced into
the UHV chamber where the substrates were cleaned by annealing at
$650^{\circ}C$ for $2$ hours in 1x10$^{-6}$ mbar oxygen to remove
hydrocarbon contaminations. Clean and well-ordered surfaces, as
determined by XPS, RHEED and LEED were obtained. Vanadium (V)
(purity $99.99\%$) was evaporated using an electron-beam
evaporator (Omicron EFM3). The deposition rate of V, measured by
moving a quartz crystal at the position of the substrate, was set
to $0.7-0.8$ \AA/min. The vanadium was simultaneously oxidized by
a beam of $^{18}$O$_2$ $(99.99\%)$. The oxygen was supplied
through a leak valve into a small buffer volume \cite{frans}
connected to the vacuum chamber through a $35$ cm long and $1$ cm
wide stainless steel pipe, ending at a distance of $10$ cm from
the substrate. The pressure in the buffer volume was measured with
a Baratron capacitance manometer.  The variation in buffer
pressure during deposition was less than $1\%$. {\vo} films with
different stoichiometries were grown by varying the $^{18}$O$_2$
buffer volume pressure, while keeping the V flux constant. During
evaporation the background pressure was 2x10$^{-8}$ mbar or lower.
The V and $^{18}$O$_2$ beams can be abruptly and simultaneously
stopped.  After closing the $^{18}$O$_2$ valve the background
pressure dropped to below 10$^{-9}$ mbar within seconds, insuring
a well-defined oxygen exposure of the sample. All the samples were
grown while keeping the substrate at room temperature (RT).

RHEED was used to monitor the evolution of the films. Thicknesses
were calibrated by monitoring the RHEED intensity oscillations
during deposition. Oscillations in the RHEED specular beam
intensity, where each oscillation corresponds to the formation of
one  new atomic monolayer (ML), allows  for control of the film
thickness. Film thicknesses were also determined \textit{ex-situ}
from X-ray specular reflectivity (XRR) measurements
\cite{mark1,mark2} and a good agreement with RHEED intensity
oscillations was obtained.

XPS measurements were performed using nonmonocromatic Al K$\alpha$
radiation ($h{\nu}=486.6$ eV) and the total energy resolution of
the electron analyzer in combination with the photon source is
about $1$ eV. To avoid {\it ex-situ} post-oxidation, a thin MgO
cap layer ($20$\AA) was grown for protection by Mg deposition in
$^{18}$O$_2$ atmosphere of 1x10$^{-8}$ mbar at room temperature.
Consequently all the {\it ex-situ} measurements, i.e. RBS, XRD,
XAS and electrical measurements were done on capped VO$_x$
samples.

To determine the oxygen content, RBS measurements have been
performed. For calibration purposes a V$_2$O$_3$ epitaxial thin
film was also grown on an Al$_2$O$_3$(0001) substrate, with the
oxygen partial pressure and the substrate temperature set at
10$^{-6}$ mbar and $550^{\circ}$C respectively \cite{horn, horn1,
horn2}. All the measurements were done in a so-called 'random'
orientation with respect to the crystalline axes, using a He$^{+}$
beam with $1.5$ MeV energy. The backscattering yields were
detected at an angle of $165$ degrees.

The XRD measurements were done using a Philips MRD diffractometer,
equipped with a hybrid mirror/monochromator for Cu K$_{\alpha}$
radiation, a $4$-circle goniometer and a programmable slit in
front of the detector.

The X-ray absorption spectroscopy (XAS) measurements were
performed at the $11$A Dragon beamline \cite{ct1,ct2} at the
Synchrotron Radiation Research Center Taiwan, using the total
electron yield mode. The light has a degree of linear polarization
of $98\%$ and an energy resolution of $0.16$ eV for photon
energies between $500$ and $550$ eV. The spectra are normalized to
the photon flux measured using a gold mesh. The angle of the
incident light is normal to the VO$_x$ films.

Resistivity measurements were performed in a commercial Quantum
design PPMS system. The electrical measurements were done in the
constant voltage mode and the current was measured along the
$[100]$ direction of the films by two and four point geometry.
Electrical contacts consisting of $10$ nm of Cr metal were
evaporated {\it ex-situ} on the MgO substrates prior to
introduction into the MBE system, in order to measure the
resistance of as-grown {\vo} films.

\section{ Results }

\subsection{Growth of {\vo} thin films}

For a wide range of incident oxygen fluxes, epitaxial growth of
{\vo} thin films on MgO(100) substrates was concluded from {\it in
situ} RHEED and LEED analysis. Our MBE system geometry allows for
monitoring changes in surface structure during film growth by
RHEED. This low angle electron diffraction technique is
particularly suited to thin film growth as it is highly sensitive
to surface morphology. A RHEED pattern of a clean MgO(100) surface
and $20${\AA} thick {\vo} thin film deposited on MgO(100) is shown
in Fig.1.

\begin{figure}
  \includegraphics[width=8cm]{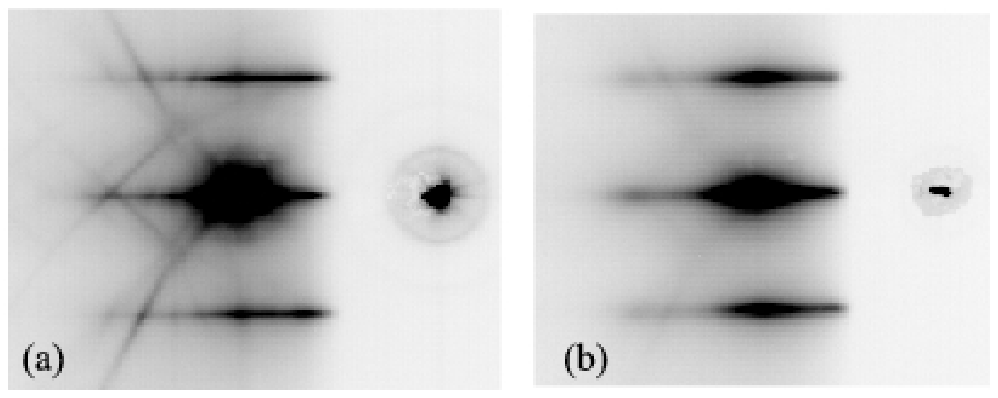}\\
  \caption{RHEED patterns recorded at an electron energy of $15$ kV
with the beam incident along $[100]$ direction. (a) clean MgO(100)
substrate.(b) $20${\AA} {\vo}(100) thin film grown on MgO(100)
substrate using $0.18$ mTorr oxygen buffer pressure while keeping
the substrate at room temperature. The picture is inverted in
order to observe better the Kikuchi lines.}
\end{figure}

The beam was incident along the $[100]$ direction. The basic RHEED
pattern did not change during growth, suggesting that the film
continues to grow as a rocksalt phase on top of the underlying
substrate. Sharp streaks and the presence of Kikuchi lines
indicate that the surface is still smooth after deposition of $10$
monolayers (ML) of {\vo}.

In most cases oscillations in the intensity of the specularly
reflected beam were observed during growth. These oscillations are
characteristic for a {\it layer-by-layer} growth mode, each
oscillation corresponding to the formation of one atomic layer.
Note that RHEED oscillations were observed at room temperature,
which is considered to be a very low temperature for an oxide
system. In Fig.2 an example of these growth oscillations is shown.

\begin{figure}
  \includegraphics[width=7cm]{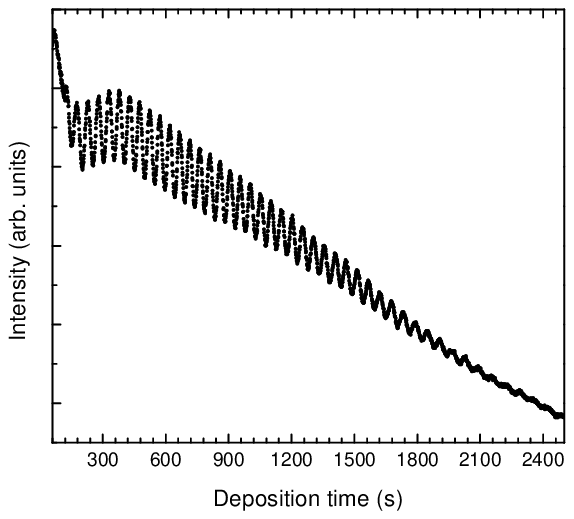}\\
  \caption{RHEED intensity oscillations of the specularly reflected
electron beam observed during deposition of {\vo} on MgO(100). The
electron beam was incident along the $[100]$ direction, with a
primary energy of $15$ kV.}
\end{figure}

In Fig.3 a sequence of RHEED patterns which were  taken after
deposition of about $50$ monolayers of {\vo} using different
oxygen fluxes is presented.
\begin{figure}[t]
  \includegraphics[width=7cm]{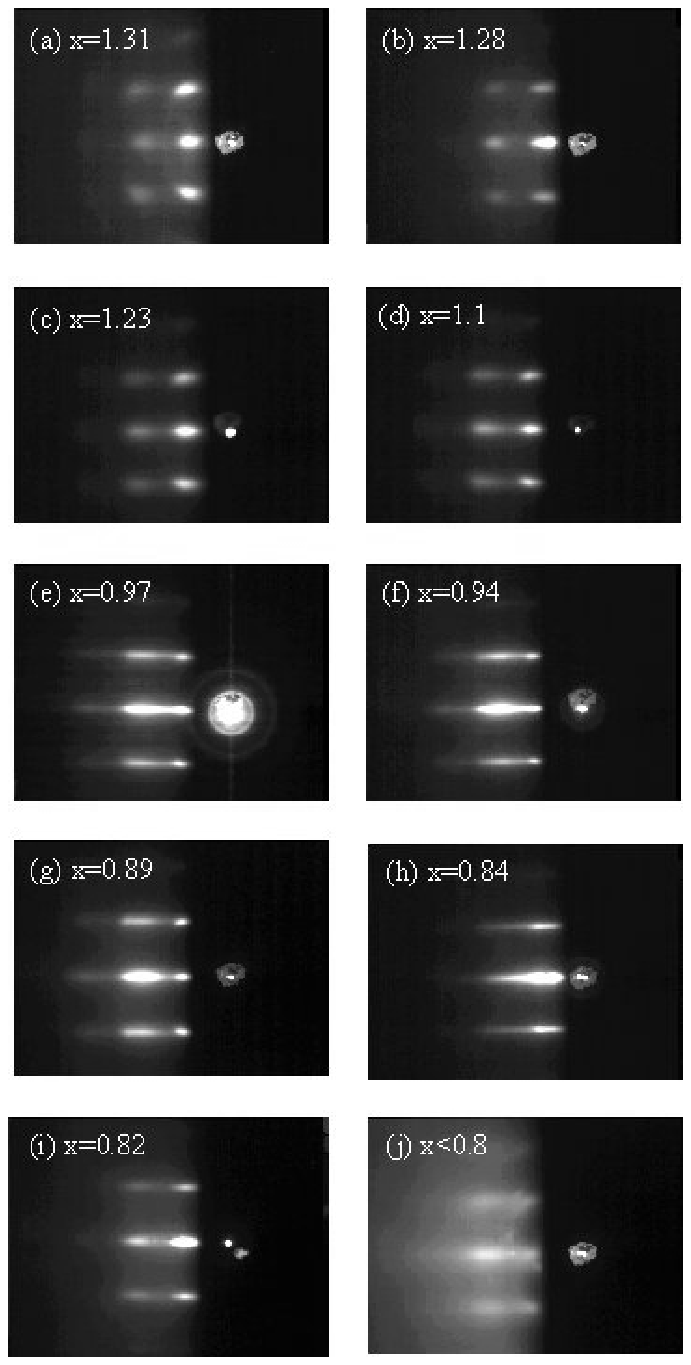}\\
  \caption{Sequence of RHEED patterns observed after deposition of
$50$ ML {\vo} on MgO(100) at different oxygen fluxes while keeping
the substrate at room temperature. The oxygen buffer pressure was
varied between $0.10$ and $0.30$ mTorr: (a) $0.30$ mTorr, (b)
$0.27$ mTorr, (c) $0.24$ mTorr (d) $0.22$ mTorr, (e) $0.20$ mTorr,
(f) $0.18$ mTorr, (g) $0.16$ mTorr , (h) $0.14$ mTorr, (i) $0.12$
mTorr, (j) $0.10$ mTorr. The electron beam was incident along
$[100]$ direction, with a primary energy of $15$ kV.}
\end{figure}

 The oxygen content $x$ is indicated on
each RHEED pattern. The corresponding oxygen flux used can be
found in the caption of Fig.3. In paragraph III.B we will explain
how the $x$ values were determined using RBS spectrometry. Varying
the oxygen buffer pressure between $0.12$ and $0.20$ mTorr
($0.82<x<0.97$) epitaxial growth is obtained and sharp streaks can
still be observed after deposition of $50$ ML {\vo}. The distance
between {\vo} streaks is the same as between the MgO streaks. This
is consistent with a coherent growth as determined {\it ex-situ}
by XRD, which will be discussed in section III.C. The rocksalt
structure of {\vo} thin films was also confirmed by LEED. The LEED
pattern displays the same square symmetry and periodicity as a
MgO$(100)$ surface.

Between $0.22$ and $0.30$ mTorr ($1.1<x<1.31$), MgO RHEED streaks
fade away quickly after starting the growth and 3D-transmission
spots appear, suggesting considerable roughening of the surface.
The absence of a LEED pattern for the $x=1.28$ and $x=1.31$ cases
confirms that the surface morphology is becoming increasingly
disordered. Nevertheless, the presence of RHEED oscillations shows
that the growth is still {\it layer-by-layer} like.

For an oxygen buffer pressure of less than $0.12$ mTorr ($x<0.8$)
powder rings and extra spots were observed in RHEED. Vanadium
atoms were not completely oxidized in the later case as deduced
from XPS core level intensities.

\begin{figure}
  \includegraphics[width=8cm]{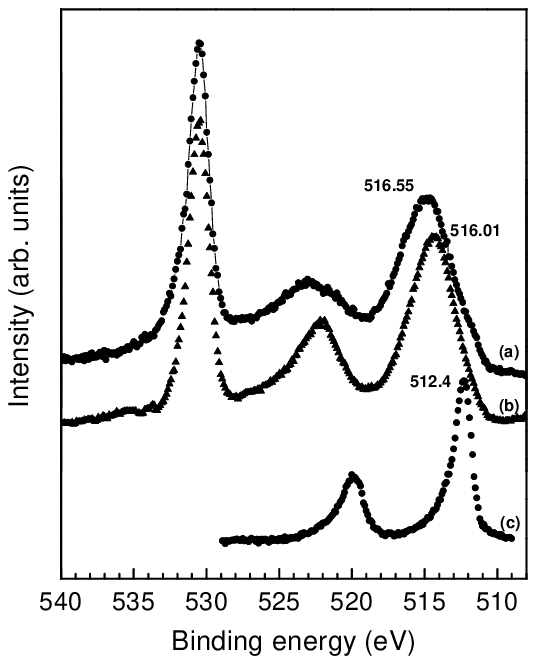}\\
  \caption{V $2p$ and O $1s$ XPS spectra of (a) V$_2$O$_3$ on
Al$_2$O$_3$ (0001), (b) {\vo} on MgO(001) and (c) vanadium metal
on MgO(001).}
\end{figure}

Attempts to make {\vo} films at higher substrate temperature were
unsuccessful. The RHEED pattern quickly disappeared or was getting
very diffuse from the beginning.

The V-$2p$ and O-$1s$ core lines measured for three different
films i.e vanadium metal on MgO$(100)$, {\vo} on MgO$(100)$ and
V$_2$O$_3$ on Al$_2$O$_3$(0001) are shown in Fig.4. The binding
energies were corrected for the charging effect by assuming a
constant binding energy of the O $1s$ peak at $531$ eV. All
spectra were corrected in a standard manner for the satellites due
to the K$\alpha$$_3$$\alpha$$_4$ components of the incident X-ray
and an integral background was subtracted afterwards. The binding
energies of the  V $2p_{3/2}$ peak found are given in the figure.
The binding energy corresponding to the V $2p_{3/2}$ in the {\vo}
film was found between the ones corresponding to V metal and
V$_2$O$_3$, respectively (see Fig.4). The vanadium oxidation state
can be in principle determined from the position and shape of the
V $2p_{3/2}$ feature. Nevertheless, due to the fact that vanadium
with different oxidation states has rather similar values for the
binding energies \cite{post} we did not perform a quantitative
analysis of the XPS data. In order to determine precisely the
stoichiometry we have adopted a different route, which will be
presented in the following section.

\begin{figure}
  \includegraphics[width=9cm]{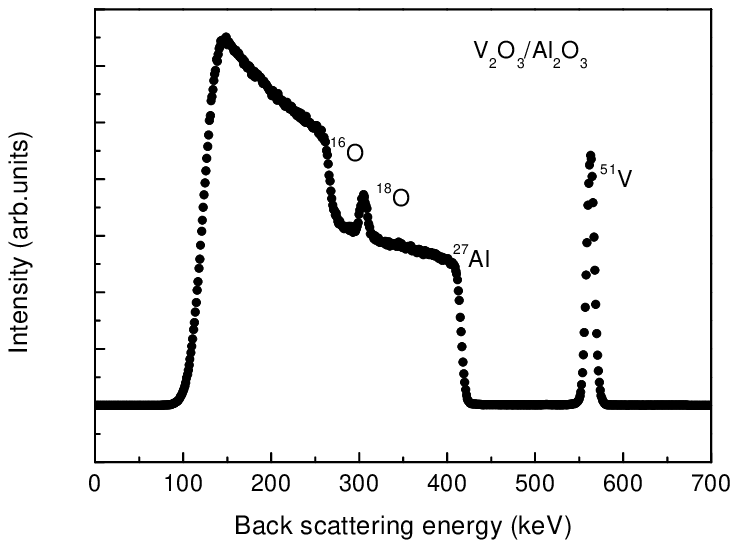}\\
  \caption{RBS spectra of $1.5$ MeV He$^{+}$ ions scattered from a
V$_2$O$_3$ film epitaxial grown on an Al$_2$O$_3$(0001) substrate.
The $^{18}$O peak can be clearly distinguished.}
\end{figure}

\begin{figure}
  \includegraphics[width=9cm]{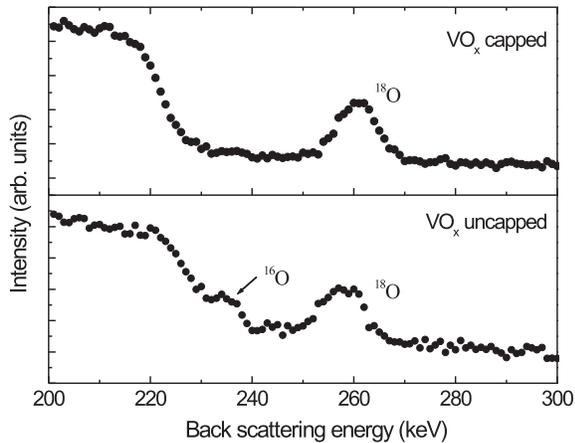}\\
  \caption{RBS spectra of $1.5$ MeV He$^+$ ions scattered from {\vo}
film deposited on MgO(100). (a) with a thin ($2$nm) MgO caplayer,
(b) without caplayer. An extra shoulder at the $^{16}$O position
is observed due to postoxidation.}
\end{figure}

\subsection{Stoichiometry determination}

The oxygen content of the films was determined from RBS
measurements. As was mentioned already in the experimental part,
$^{18}O_{2}$ instead of $^{16}O_{2}$ was employed for film growth
to distinguish between the oxygen of the film and substrate. Using
a 1.5 MeV He$^{+}$ beam, a good mass separation between $^{16}$O
from the substrate and $^{18}$O from the film can be obtained.
This can be observed in Fig.5 showing a RBS spectrum of a
V$_{2}$O$_{3}$ layer grown epitaxially on an Al$_{2}$O$_{3}$(0001)
substrate, which was used for calibration. A nice hexagonal LEED
pattern characteristic for the corundum structure was observed,
proving the high quality and long range order of the
V$_{2}$O$_{3}$ film. In Fig.6 part of the spectra for a capped and
uncapped {\vo} layer on MgO(001) is shown. In both spectra a
well-separated $^{18}$O-peak is visible. For the uncapped layer,
an additional shoulder at the $^{16}$O position can be observed
due to the fact that the layer is post-oxidized in air. The
absence of this shoulder in the spectrum of the capped layer
proves that capped {\vo} films are really protected from
post-oxidation.

\begin{figure}
  \includegraphics[width=9cm]{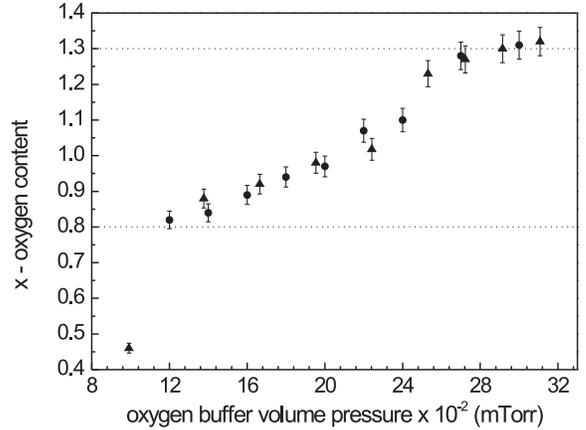}\\
  \caption{The oxygen content $x$ as a function of the buffer
volume pressure as determined from RBS data for two series of
{\vo} samples.}
\end{figure}

With the following formula we have calculated the ratio between
the numbers of V and O atoms in a VO$_{x}$ film, using
V$_{2}$O$_{3}$ as a reference sample.

\begin{equation}
 x=\frac{3}{2}\frac{\lbrack\frac{A_O}{A_V}\rbrack_{_{VO_x}}}{\lbrack\frac{A_O}{A_V}\rbrack_{_{V_2O_3}}}
\end{equation}

The areas A$_O$ and A$_V$ in the RBS spectra were determined by
fitting the V and $^{18}$O peaks from VO$_{x}$ and V$_{2}$O$_{3}$
samples to a gaussian function, in the case of the $^{18}$O-peak
after subtraction of the linear background due to Mg. The
$^{18}$O-peak area was corrected for the contribution of the
caplayer, using the thickness of the {\vo} films and the
Mg$^{18}$O caplayer, as determined from the RHEED oscillations.
The resulting $x$-values are plotted in Fig.7 as a function of the
oxygen buffer volume pressure for two series of samples. One can
observe that the $x$ values are nicely reproducible. The error in
the determination is about $3\%$. The upper and lower
stoichiometry limit found are similar to the ones known for bulk
material. For the sample grown at $0.10$ mTorr the vanadium metal
is not completely oxidized, as deduced from the XPS spectra. For
$x$ bigger than $1.3$ a mixture of two phases was found.

\subsection{ Structure of {\vo} thin films }

From RHEED and LEED patterns it was already concluded that {\vo}
films grow epitaxially on MgO(100). To analyze the epitaxy and
crystal structure of the films in a more quantitative way, we also
performed an {\it ex-situ} XRD analysis.

The measurements were done on samples capped with a thin
($20${\AA}) epitaxial MgO film. $\theta$-$2$$\theta$ scans show
only weak diffraction peaks close to the $(002)$ and $(004)$ peaks
of the MgO substrate, as expected from films having a rocksalt
structure. The reflections are broadened due to the finite
thickness of the film. There are no peaks corresponding to other
phases. Moreover, for samples made with the oxygen buffer pressure
varying between $0.12$ and $0.22$ mTorr a number of subsidiary
thickness fringes were observed, suggesting a well defined
composition and thickness of the film (see Fig.8).

To determine whether the growth is fully coherent or partially
relaxed, the most convenient way is to measure the intensity
profile around a non-specular reflection common to the substrate
and the overlayer. This can be done by performing  a set of
$2\theta$-$\omega$ scans at different $\omega$, $2\theta$ and
$\omega$ being the detector and sample orientations with respect
to the beam direction. These scans can be easily mapped into
reciprocal space. An example of such a map for the region in
reciprocal space around the $(113)$ reflection is shown in Fig.9.
The horizontal and vertical axes are k vectors parallel
($k_{par}$) and perpendicular ($k_{per}$) to the surface plane,
respectively. The intensity scale in the figure is logarithmic.
The feature corresponding to the {\vo} film can be clearly
distinguished. The elongated shape of the MgO reflection
perpendicular to the radial direction is due to mosaic spread. The
peaks of MgO and {\vo} are at the same $k_{par}$ value, which
proves that the film is fully coherent. Consequently, {\vo} thin
films experience a compressive strain in the perpendicular
direction, which is induced by the lattice mismatch.

\begin{figure}
  \includegraphics[width=7cm]{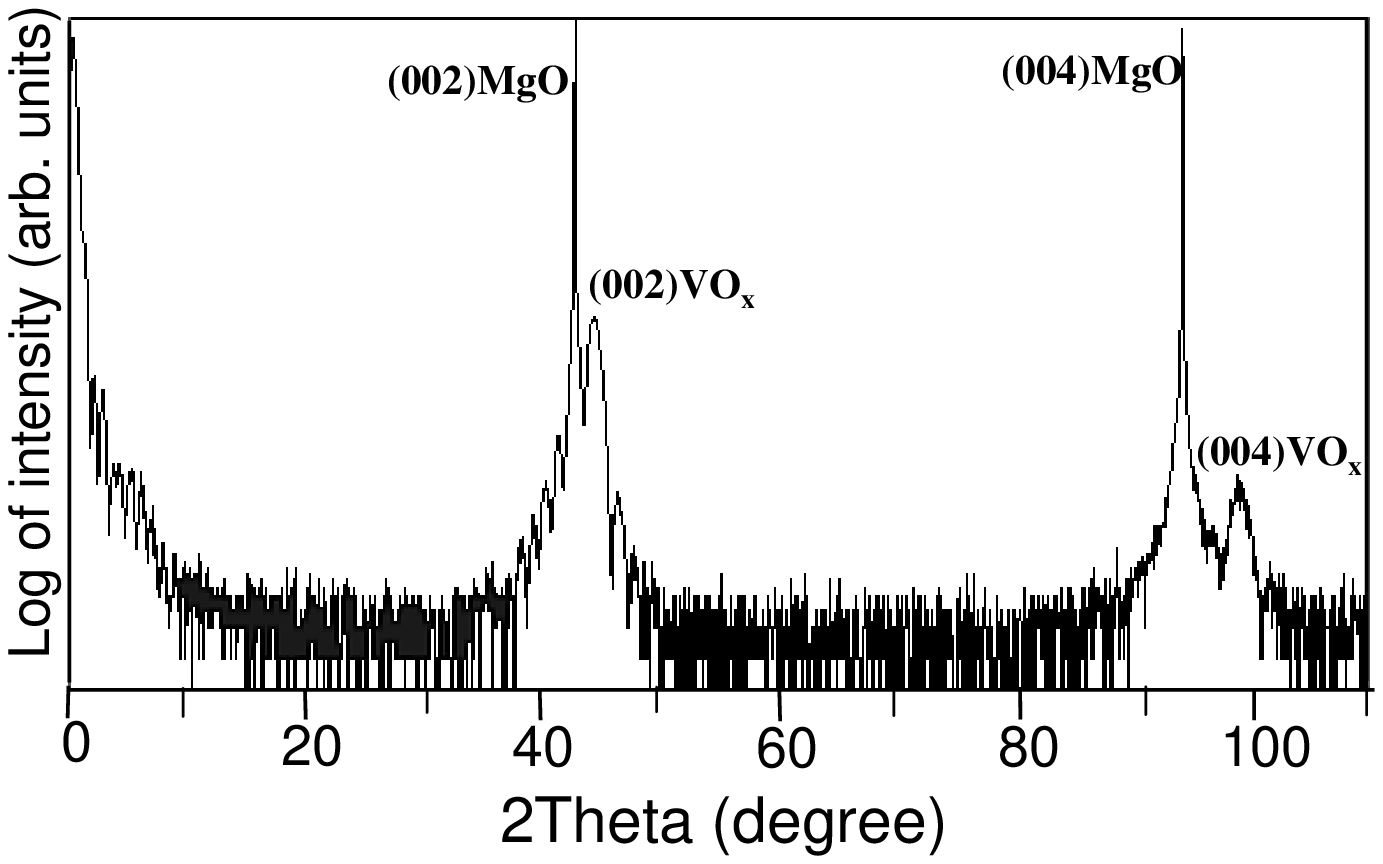}\\
  \caption{$\theta$-$2$$\theta$ X-ray diffraction measurement of
$10$nm thick {\vo} thin film prepared with $x=0.94$. Only the
$(002)$ and $(004)$ diffraction peaks characteristic of the
rocksalt structure can be observed. The reflections from the film
are broadened due to the finite thickness of the film. Subsidiary
thickness fringes indicate a well defined composition and
thickness of the film.}
\end{figure}

\begin{figure}
  \includegraphics[width=6cm]{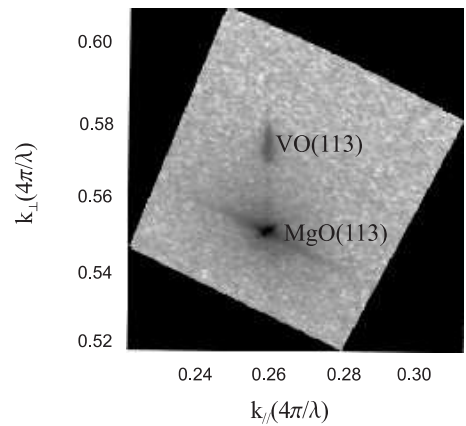}\\
  \caption{XRD map of reciprocal space around the non-specular
$(113)$ reflection of a $10$nm thick {\vo} thin film with
$x=0.94$. The logarithm of the diffracted intensity as a function
of the in-plane $k_{\parallel}$ and out-of-plane $k_{\perp}$
reciprocal lattice vectors is plotted. The $x$- and $y$- axis are
in units of $\frac{4\pi}{\lambda}$, with $\lambda = 0.15015$ nm.}
\end{figure}
The lattice constant normal to the substrate surface can be
determined from $\theta-2\theta$ scans around MgO(002) or MgO(004)
peaks using the Bragg law. 2$\theta$ values were determined by
fitting the peaks to a gaussian function. The perpendicular
lattice constant is plotted in Fig.10 as a function of the oxygen
content.

\begin{figure}
  \includegraphics[width=9cm]{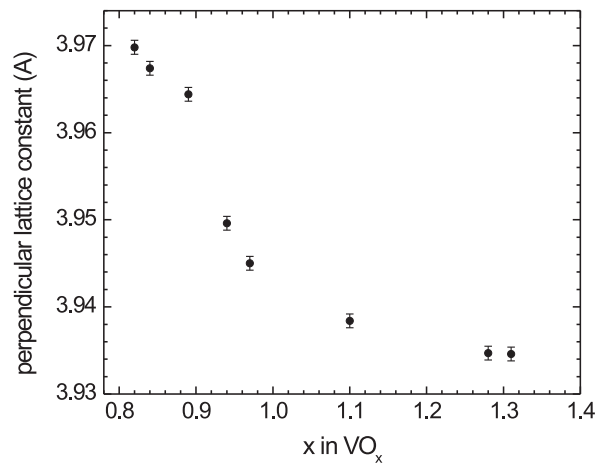}\\
  \caption{Perpendicular lattice constant of {\vo} films with
different oxygen content as a function of $x$ as determined from
XRD. The perpendicular lattice constant is decreasing with
increasing $x$.}
\end{figure}

 We found that the perpendicular lattice constant and
consequently also the average lattice constant is decreasing with
increasing oxygen content. This is in contrast to the behavior
reported by Banus \cite{banus} for bulk {\vo}.

The layer thicknesses determined from X-ray specular reflectivity
(XRR) measurements were in good agreement with the values obtained
from the RHEED intensity oscillation period.

\subsection{ Vacancy concentration }

As was already mentioned in the Introduction, a large number of
both vanadium and oxygen vacancies is characteristic for bulk
{\vo}. Using a combination of RHEED and RBS resuls, an estimation
of the number of vacancies present in {\vo} thin films can be
made. RHEED oscillation periods were determined for all the {\vo}
samples as well as for a calibration V$_2$O$_3$ thin film on
Al$_2$O$_3$(0001). All samples were grown using the same value for
the vanadium flux. By comparing the time needed to grow one
monolayer of {\vo} and V$_2$O$_3$ and assuming that V$_2$O$_3$ is
stoichiometric and free of vacancies, the fraction of vacant
vanadium sites ($V_V$) in {\vo} can be directly calculated. The
oxygen vacancy concentration ($V_O$) can be obtained because $x$
is already known from RBS, using the following expression:

\begin{equation}
 x = \frac{1-V_O}{1-V_V}
\end{equation}

As illustrated in Fig.11, the vanadium vacancy concentration is
increasing with increasing oxygen content $x$, while the oxygen
vacancy concentration is decreasing. The total number of vacancies
(V$_V$+V$_O$) is also decreasing with $x$. Close to the
stoichiometric value we found about 16 {\%} vacant sites of both
vanadium and oxygen.

\begin{figure}
  \includegraphics[width=9cm]{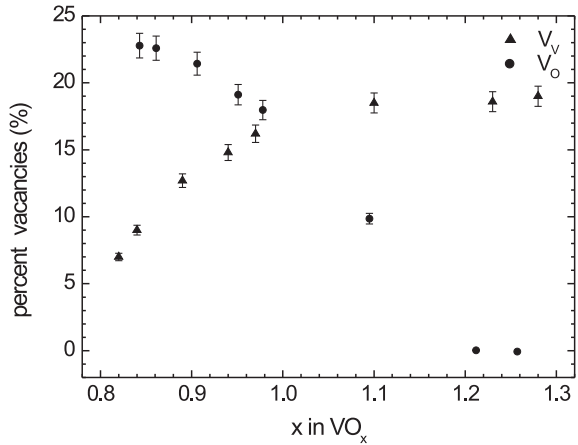}\\
  \caption{Vanadium and oxygen vacancy concentrations in {\vo} as
determined from the period of RHEED oscillations. The number of
vanadium vacancies is increasing with $x$, while the number of
oxygen vacancies is decreasing.}
\end{figure}

\subsection{Valence of the V ions from XAS}

XAS measurements were done at the O \textit{K}- edge in order to
verify the valence of the V and O ions. The data are depicted in
Fig.12 for samples with the oxygen content $x$ (as determined from
RBS) between $0.8$ and $1.3$.  For comparison, we have also
included the spectrum of a V$_{2}$O$_{3}$ film in top part of the
figure. The O \textit{K}- edge absorption spectra correspond to
the dipole-allowed transitions from the O $1s$ to the O $2p$
shell, which is partially empty due to the hybridization with V
$3d$ conduction band states. The spectral structures that can be
observed in the $528-534$ eV photon energy range are dictated by
these V $3d$ states \cite{abbate,degroot,elp,nakai}. For photon
energies higher than $535$ eV, the oxygen of the MgO cap layer
also start to contribute to the XAS signal \cite{nakai}.

The distinct peak observed at photon energies around $532$ eV, can
be assigned to transitions into the empty V
$3d$-$e_{g}$$^\uparrow$ band \cite{elp}. Structures at higher
photon energies can be ascribed to transitions to the higher lying
V $4sp$ related bands. Important is to note that for the most
oxygen deficient VO$_{x}$ samples, i.e. $x<<1$, the lowest
spectral structure is given by the $532$ eV peak, indicating that
the lower lying V $3d$-$t_{2g}$$^\uparrow$ band is completely
occupied by three electrons, and that thus the V valence is $2+$
or less. For samples with higher oxygen content, i.e. for $x>1$, a
clear low energy peak appears at about $530$ eV. This strongly
suggests that holes are introduced in the V
$3d$-$t_{2g}$$^\uparrow$ band, meaning that the V valence is
higher than $2+$. This assignment is supported by the
V$_{2}$O$_{3}$ spectrum, in which transitions to both the
$t_{2g}$$^\uparrow$ and the $e_{g}$$^\uparrow$ bands are possible
because only two electrons occupy the three-fold degenerate
$t_{2g}$$^\uparrow$ orbital in this V$^{3+}$ $3d^2$ system.

These XAS measurements show that the cross-over from less than
$2+$ to more than $2+$ V valencies occurs for an RBS $x$ value of
about $0.94-0.97$. This value is not very far from $1.00$, and can
be taken as an indication for a good agreement between the XAS and
RBS methods.

We have also carried out XAS experiments at the V
\textit{L$_{23}$}($2p\rightarrow3d$) edges. The results are shown
in Fig.13, in which we have also included the spectra for
V$_{2}$O$_{3}$ and Cr$_{2}$O$_{3}$ for reference purposes. In
going from $x=0.8$ to $x=1.3$ we can observe a gradual change in
the spectra. Distinct and sharp structures start to develop for
$x\geq1$. The similarity of these structures with those of
V$_{2}$O$_{3}$ is striking, and in fact indicates that V ions in a
$3+$ valence state are present for $x\geq1$, consistent with the
observations at the O \textit{K} edge mentioned above.

Important is the observation that the spectra are quite broad for
$x$ around 1. In the simplest approximation, one would expect to
see a spectrum with the typical atomic multiplet structure of a
$2p^{6}3d^{3}\rightarrow2p^{5}3d^{4}$ transition of a $V^{2+}$ ion
in a $O_{H}$ symmetry, as shown in the bottom curve of Fig.13.
This curve has been calculated using standard parameters
($10D_{q}\simeq2$ eV) often applied for the analysis of transition
metal oxide soft-x-ray absorption spectra
\cite{degroot2,thole,tanaka} and it shows quite distinct features
with several peaks and valleys. Clearly, the experimental V
\textit{L$_{23}$} spectra for $x\approx1$ do not have these
distinct features.

\begin{figure}
  \includegraphics[width=9cm]{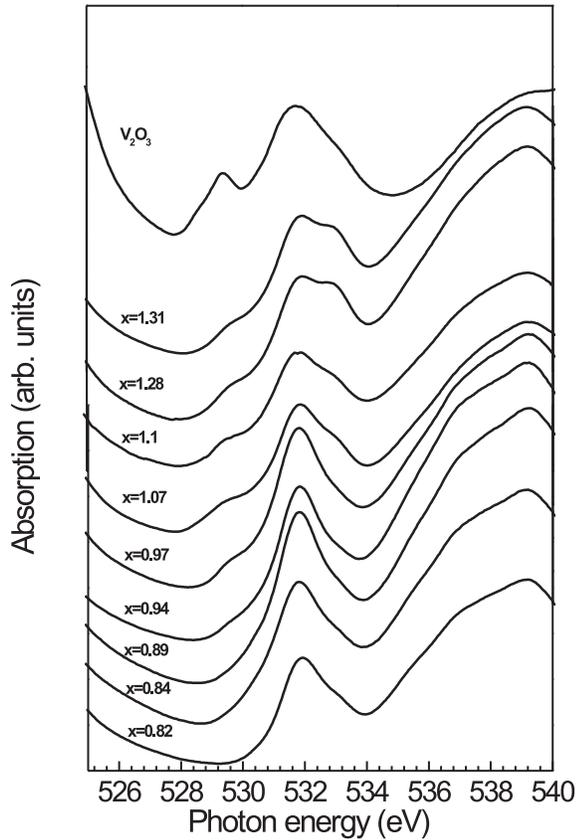}\\
  \caption{O $1s$ X-ray absorption spectra of {\vo} samples with the
oxygen content varying between $0.8$ and $1.3$.}
\end{figure}

\begin{figure}
  \includegraphics[width=9cm]{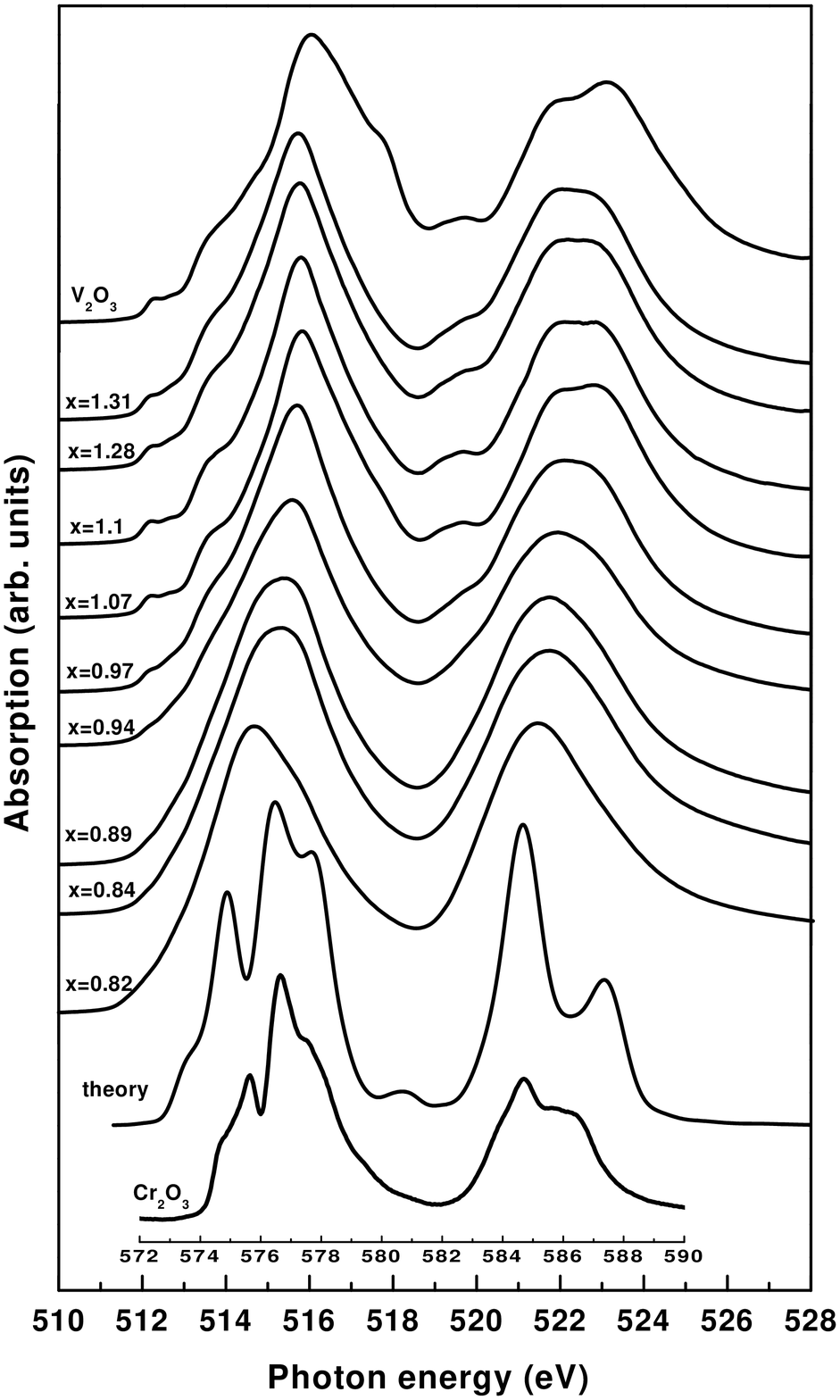}\\
  \caption{V \textit{L$_{23}$} X-ray absorption spectra of {\vo}
samples with the oxygen content varying between $0.8$ and $1.3$.}
\end{figure}

Also a comparison with the experimental spectrum of
Cr$_{2}$O$_{3}$ (see Fig.13), which is also a $3d^{3}$ ion in
approximately $O_{H}$ local symmetry, leads to the conclusion that
the V \textit{L$_{23}$} spectra are anomalously broad. We take
this observation as an indication that the vanadium in VO is not
all in the local $3d^{3}-O_{H}$ symmetry, but that instead an
appreciable amount of vanadium ions are experiencing strong local
ligand fields of low symmetry associated with the presence of
large amount of vacancies. These low symmetry ligand fields must
be at least several hundreds of meV strong, in order to wash out
completely the multiplet structure of a  $3d^{3}$-ion in $O_{H}$
symmetry. The implication of such fields will be discussed in
Section IV.

\subsection{Transport properties}

Electrical measurements have been performed to study the
electronic structure of {\vo} films. Resistivity data have been
reported previously  by Banus \cite{banus} for polycrystalline
samples. Bulk material exhibits a semiconducting behavior for $x>
1$, with an activation energy rising to about 40 meV for $x=1.3$.
For $x<1$ {\vo} behaves like TiO$_x$, with an almost temperature
and composition independent resistivity of about 3x10$^{-3}$
$\Omega$cm. Banus et al. reported a transition from a semimetallic
to semiconducting type behavior in {\vo} at $x=1.05$.

\begin{figure}
  \includegraphics[width=9cm]{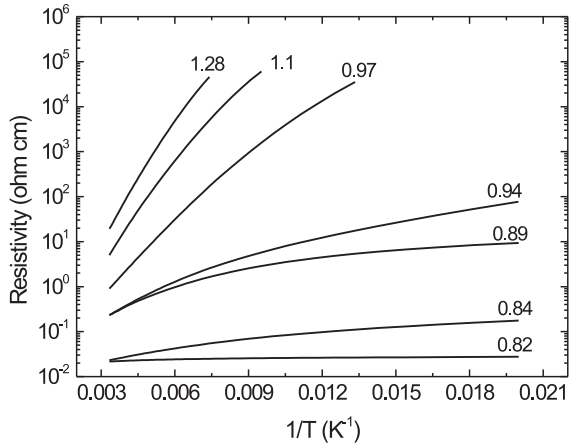}\\
  \caption{Logarithm of resistivity of {\vo} films with different
oxygen content plotted against $1/T$. Resistivity is increasing
with decreasing temperature except for the $x=0.82$ sample,
suggesting a semiconductor-like behavior.}
\end{figure}

\begin{figure}
  \includegraphics[width=9cm]{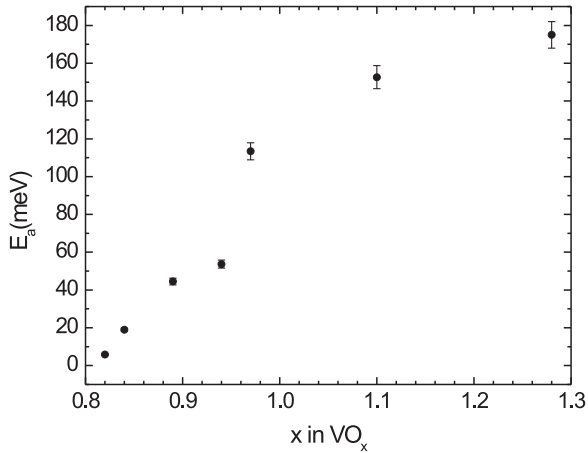}\\
  \caption{Activation energy $E_a$ versus $x$ calculated for the
temperature interval $150<T<300$ K.}
\end{figure}

The variation of the logarithm of the electrical  resistivity of
{\vo} thin films with different oxygen content is shown as a
function of $1/T$ in Fig.14. The resistivity is increasing with
decreasing T for all but the $x=0.82$ sample, suggesting a
semiconductor-like behavior in the entire temperature range
studied. However, it is evident in Fig.14 that $\log\rho$ does not
vary linearly with $1/T$. Fig.15 shows the composition dependence
of the activation energies calculated from the $\log\rho$-vs-$1/T$
plots close to room temperature. Note, that the apparent
activation energy decreases with decreasing temperature.

Such behavior is similar to that described by Banus \cite{banus},
but we did observe a few important differences. Compared to the
corresponding resistivity results for bulk samples we found  a
much higher absolute value of the resistivity. As can be observed
from Fig.14 the transition from a metallic to a semiconductor type
behavior is shifted from $x=1$, as found in bulk material, to
$x=0.8$ for thin films. Moreover, the activation energy calculated
close to room temperature (Fig.15) is larger. For example, for
bulk stoichiometric VO, Banus et al.\cite{banus} obtained  an
apparent activation energy of only $5$ meV. For thin films we
found a much higher value of $\approx$ $110$ meV. A large increase
in the activation energy can be observed going from $x=0.94$ to
$x=0.97$. Although the oxygen content is changing only with $3\%$,
the activation energy is becoming almost twice as large. For $x$
$\approx$ $1.3$ the activation energy rises to $180$ meV.

\begin{figure}
  \includegraphics[width=9cm]{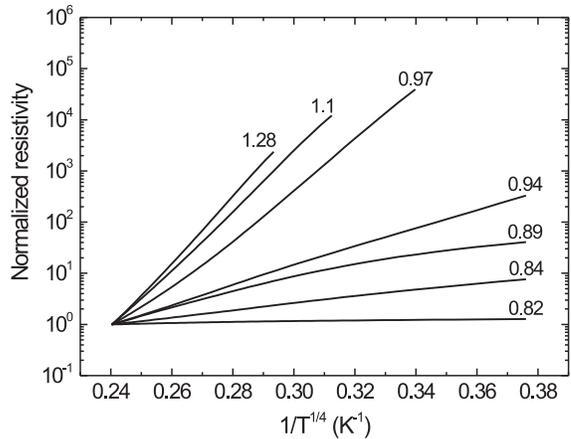}\\
  \caption{Logarithm of normalized resistivity of {\vo} thin films
with different oxygen content plotted as a function of
T$^{-1/4}$.}
\end{figure}

The T-dependence of the resistivity  is better described by
variable range hopping behavior \cite{mott1} given by:

\begin{equation}
 \rho\approx\exp(\frac{T_0}{T})^{\frac{1}{4}}
\end{equation}

This is evident from Fig.16 showing the normalized resistivity
versus $T^{-1/4}$. This result is not surprising knowing that a
high number of vacancies is characteristic for this material. A
random field caused by vacant sites can produce localization of
the electrons near the Fermi level just like in amorphous
materials. Its occurrence in {\vo} seems to be strong evidence
that the vacancies play a major role, causing large changes in its
electronic properties. The parameter $T_0$ can be extracted from
the slope of the curves in Fig.16. $T_0$ is related to the density
of localized states at the Fermi level N(E$_F$)\cite{brown} by the
following expression:

\begin{equation}
 T_0 = \frac{24\alpha^3}{{\pi}N(E_F)k_B}
\end{equation}
\begin{table}
\caption{ Parameters extracted from T$^{-1/4}$ dependence of
resistivity.}
\begin{tabular}{c|c|c}
x in {\vo}&T$_0$(K)&N(E$_F$)(cm$^{-3}$eV$^{-1}$)\\
\tableline
\\
0.82&2.6&2$\times10^{26}$\\
0.84&2$\times10^3$&2$\times10^{23}$\\
0.89&5$\times10^4$&8$\times10^{21}$\\
0.94&12$\times10^4$&3$\times10^{21}$\\
0.97&5$\times10^6$&1$\times10^{20}$\\
1.1&1$\times10^7$&4$\times10^{19}$\\
1.28&2$\times10^7$&2$\times10^{19}$\\
\end{tabular}
\end{table}
where $\alpha$$^{-1}$ is the decay length of the wave function
associated with the charge carriers and $k_B$ is the Boltzmann
constant. Taking a reasonable value for $\alpha$$^{-1}$ as
$5${\AA}, N(E$_F$) can be calculated from the values of $T_0$.
These parameters are tabulated in Table 1 together with the
corresponding $x$ values. Clearly, one can observe that for
$x=0.82$ sample the density of localized states at the Fermi level
is high and  starts to decrease with increasing the oxygen
content.

\section{Discussion}

The transport properties of VO$_x$ can not be understood on the
basis of a simple band picture, because the material should have
been metallic. Shortly after the publication of Banus et al., two
papers by Goodenough \cite{goodenough} and Mott \cite{mott}
appeared, in which explanations were given in terms of electron
correlation effects and vacancies. Both authors assume a finite
Hubbard U, which is large enough to split the $t_{2g}$ band.
Because V$^{2+}$ supplies $3$ electrons to the $t_{2g}$ band, the
valence band of VO is given by a completely filled lower Hubbard
band, separated by a gap from the conduction band which is the
empty upper Hubbard band.

According to Mott \cite{mott}, the random field produced by the
vacancies causes Anderson localization \cite{anderson} in the
overlap region of the two Hubbard bands. This explains why the
conductivity in {\vo} is of the variable range hopping type at low
temperatures. In Goodenough's model \cite{goodenough} the effect
of the vacancies on the electronic structure is discussed much
more explicitly. He assumes a high degree of trapping of electron
and holes near anion and cation vacancies, respectevely. In this
way the Fermi-level stays near the minimum in the DOS of {\vo},
explaining its semiconductor-like behavior. The transition to a
more itinerant behavior for $x<1$ is attributed to a tail in the
trapped electron distribution. A second effect of the trapping of
charges near vacancies is that the loss of Madelung energy is
limited.

The presence of a large number of vacancies in {\vo} remains one
of the most puzzling characteristics. The only two other binary
oxides showing this behavior are the rocksalts TiO$_x$ and NbO,
with NbO having about $25\%$ vacancies \cite{neckel1}. Apparently
the creation of vacancies stabilizes the crystal by reducing the
Gibbs free energy. According to Goodenough \cite{goodenough} the
formation of vacancies leads to a reduction in the lattice
constant, thereby broadening the $t_{2g}$ bands. The resulting
stabilization of the occupied valence band states is counteracted
by a reduction in Madelung energy, but this energy loss is assumed
to be minimized by a localization  of the charge compensating
electrons and holes near the oxygen and vanadium vacancies
respectively. He argues that the application of hydrostatic
pressure should reduce the vacancy concentration considerably and
he refers to experiments giving indirect evidence in favor of this
effect \cite{banus2}.

The {\vo} films we have grown epitaxially on MgO substrates are
thin enough to be coherent with the substrate and are under
tensile stress. In fact, the in-plane lattice constant is expanded
($\approx$3$\%$), while the out-of-plane lattice constant is
reduced due to the Poisson effect. Applying Goodenough's vacancy
induced lattice contraction arguments, we would expect that the
formation of vacancies will not be less favored in our thin film,
because the lattice constant of the film is fixed by the
substrate, thus preventing the energy gain to occur that otherwise
is associated with the lattice contraction and broadening of the
$t_{2g}$ bands. In contrast to this expectation, however, we found
that both the cation and anion concentration are not much
affected. For stoichiometric VO film, the values are even slightly
higher (of the order of $1$ or $2\%$) than for bulk material.

We agree with Goodenough, nevertheless, that the mechanism for
vacancy formation must be searched for in terms of energy
arguments. To our knowledge, \textit{ab-initio} total energy
calculations have not yet been carried out to study the stability
of vacancy formation in this material. We may speculate about the
mechanism by reviewing Goodenough's arguments \cite{goodenough}.
Goodenough starts with the picture that electrons and holes are
being trapped near the cation and anion vacancies, as to avoid
part of the Madelung energy loss. The $t_{2g}$ orbitals of cations
neighboring a cation vacancy are then destabilized by a reduced
bonding to neighboring $t_{2g}$ orbitals and a stronger $\pi$
bonding to the $p$ orbitals next to the vacancy, thereby raising
their energy above the Fermi level. On the other hand, near an
oxygen vacancy the bonding $t_{2g}$ orbitals are stabilized,
whereas the $e_{g}$ and $s$ orbitals are less destabilized by the
absence of oxygen $p_{\sigma}$ orbitals at the vacant site. These
effects contribute to the stabilization energy of vacancies, but
not enough. Instead of looking for a possible further energy gain
in the $t_{2g}$ band due to lattice contraction induced band
broadening, we now discuss what low symmetry ligand fields can do
to lower the energy of the occupied $t_{2g}$, motivated by the
fact that the XAS measurements reveal that those ligand fields are
very strong in VO$_x$, i.e. with a strength of the order of
several hundreds meV.

We will first discuss the case of oxygen vacancies. Neighboring
divalent $3d$ metal ions are surrounded by five oxygen ligands in
a square pyramidal arrangement with C$_{4v}$ symmetry. The
$t_{2g}$ and $e_g$ levels will be split into three non- and one
doubly degenerate levels, respectively. With two electrons being
trapped at or near the vacancy, we have one extreme possibility in
that both electrons are trapped at the vacant site, forming an
$F$-center, and the other possibility in that the excess electrons
are trapped at neighboring $d$-metal sites. Assuming that at least
a fraction of the excess electrons resides at the $d$-metal ions,
the average number of $d$ electrons will be larger than three for
V-ions next to a vacancy. In both cases some ligand field
stabilization energy is gained, owing to the electrons occupying
the lower $e_g$ level in VO$_{x<1}$.

Considering now the case of $d$ metal vacancies, we may speculate
that two excess holes are trapped near these vacancies in
VO$_{x>1}$. One extreme possibility is that the holes are residing
at oxygen neighbors. Following a suggestion of Elfimov et al.
\cite{elfimov}, this would give a stable configuration in which
the holes are in a triplet state localized at the oxygen
coordination polyhedron around the vacancy. However, in contrast
to the case of CaO discussed by these authors, in a transition
metal compound there is also the possibility that the two holes
are located at next nearest neighbor transition metal ions. The
presence of the cation vacancy leads to a local symmetry lowering
(C$_{2v}$) at these cation sites. Also in this case a ligand field
stabilization is anticipated, because the average number of $d$
electrons on $d$-metal ions next to a vacancy will be less than
three for V.

Although, as was argued above, the decrease of direct overlap
between $t_{2g}$ orbitals in the strained films does not have a
major effect on the concentration of vacancies, it will change the
band structure in the valence band region and consequently change
the electrical properties. Expanding the lattice by epitaxial
growth on a substrate with a larger lattice constant like MgO will
decrease the bandwidth and therefore increase the gap. This
explains the lower conductivity and higher room temperature
activation energies found in our films. In the models proposed by
Goodenough \cite{goodenough} and Mott \cite{mott}, the on-site
Coulomb interaction U is assumed to open up a small energy gap in
the itinerant $t_{2g}$ band. It is assumed that overlapping tails
of localized states, which are related to the disorder associated
with the large number of defects in the system, are present at the
edges of these Hubbard bands. In this way Mott \cite{mott}
explains why in bulk samples the conductivity vs temperature is
semiconductor-like at high temperature, but is better described by
a variable range hopping mechanism at low temperatures. The
conductivity results for our films are consistent with this
explanation. The main difference is that the resistivity and
apparent activation energies of the films are much higher, and
that the transition from semiconducting to metallic behavior is
shifted to lower $x$ values, i.e. from $x\approx1$ in bulk samples
to $x\approx0.8$ in films.

\section{Conclusions}

We have successfully grown epitaxial {\vo} films on MgO(100)
substrates. Up to at least 120{\AA} the growth is coherent and
layer-by-layer-like. $^{18}$O$_{2}$-RBS was introduced as a
convenient method to determine accurately the stoichiometry of
these ultrathin layers. Once the stoichiometry is known, the
vacancy concentration of both vanadium and oxygen can be
calculated from the time to grow one monolayer as determined from
RHEED. The numbers turn out to be very similar to those for the
bulk material. This implies that the formation of a high
concentration of vacancies may not be directly related to an
increased $t_{2g}$ band width as a result of a vacancy induced
lattice contraction. Instead we suggest that a detailed study is
required to calculate the possible stabilization due to additional
ligand field splittings at the low symmetry metal sites near the
vacancies, as observed from XAS. Nevertheless, the decrease in
direct overlap between $t_{2g}$ orbitals and the concomitant
increase of the size of the pseudo-gap between the lower and upper
$t_{2g}$ Hubbard bands is held responsible for the much larger
electrical resistivity in strained {\vo} films.

\section{Acknowledgements}
We would like to thank D. I. Khomskii, T. T. M. Palstra and D. O.
Boerma for stimulating discussions, as well as H. Bruinenberg and
J. Baas for skillful technical assistance. We would like to thank
A. Tanaka for the use of the code to calculate the XAS spectra.
The research of M.W.H. and L.H.T. is supported by the Deutsche
Forschungsgemeinschaft through SFB 608.

\end{document}